\documentclass[a4paper,fleqn,usenatbib]{mnras}
\usepackage[T1]{fontenc}
\usepackage{ae,aecompl}
\usepackage{graphicx}
\usepackage{amsmath}
\usepackage{amssymb}

\newcommand{\kms}{\,km\,s$^{-1}$}

\usepackage{longtable}

\title[Line Identification] {Line identification
and photometric history of the hot post-AGB star Hen~3-1013
(IRAS~14331-6435)}
\author[Arkhipova et al.]{V. P. Arkhipova$^{1}$,
M. Parthasarathy$^{2}$\thanks{E-mail:m-partha@hotmail.com},
N. P. Ikonnikova$^{1}$, M. Ishigaki$^{3}$, \newauthor
S. Hubrig$^{4}$, G. Sarkar$^{5}$,
A. Y. Kniazev$^{1,6,7,8}$\\
$^{1}$ Lomonosov Moscow State University, Sternberg Astronomical
Institute, 13 Universitetskij prospekt, Moscow 119234, Russia\\
$^{2}$ Indian Institute of Astrophysics, Bangalore 560034, India\\
$^{3}$ IPMU, The University of Tokyo, Japan\\
$^{4}$ Leibniz Institute for Astrophysics (AIP), Potsdam,
Germany\\
$^{5}$ Department of Physics, IIT Kanpur, India\\
$^{6}$ South African Astronomical Observatory, Cape Town, South
Africa\\
$^{7}$ Southern African Large Telescope Foundation, Cape Town,
South Africa\\
$^8$ Special Astrophysical Observatory of RAS, Nizhnij Arkhyz,
Karachai-Circassia 369167, Russia}

\date{Accepted 2018. Received 2018; in original form 2018}

\pubyear{2018}

\begin{document}
\label{firstpage}
\pagerange{\pageref{firstpage}--\pageref{lastpage}} \maketitle

\begin{abstract}

We present a study of the high-resolution optical spectrum for the
hot post-asymptotic giant branch (post-AGB) star, Hen~3-1013,
identified as the optical counterpart of the infrared source
IRAS~14331-6435. For the first time the detailed identifications
of the observed absorption and emission features in the wavelength
range 3700-9000 \AA\ is carried out. Absorption lines of
\ion{H}{i}, \ion{He}{i}, \ion{C}{I}, \ion{N}{I}, \ion{O}{I},
\ion{Ne}{I} \ion{C}{II}, \ion{N}{II}, \ion{O}{II}, \ion{Si}{II},
\ion{S}{II}, \ion{Ar}{II}, \ion{Fe}{II}, \ion{Mn}{II},
\ion{Cr}{II}, \ion{Ti}{II}, \ion{Co}{II}, \ion{Ni}{II},
\ion{S}{III}, \ion{Fe}{III} and \ion{S}{IV} were detected. From
the absorption lines, we derived heliocentric radial velocities of
$V_\text{r}=-29.6\pm0.4$ {\kms}. We have identified emission
permitted lines of \ion{O}{I}, \ion{N}{I}, \ion{Fe}{II},
\ion{Mg}{II}, \ion{Si}{II} and \ion{Al}{II}. The forbidden lines
of [\ion{N}{I}], [\ion{Fe}{II}], [\ion{Cr}{II}] and [\ion{Ni}{II}]
have been identified also. Analysis of [\ion{Ni}{II}] lines in the
gaseous shell gives an estimate for the electron density
$N_e\sim10^7$ cm$^{-3}$ and the expansion velocity of the nebula
$V_\text{exp}=12$ \kms. The mean radial velocity as measured from
emission features of the envelope is $V_\text{r}=-36.0\pm0.4$
{\kms}. The Balmer lines \ion{H}{$\alpha$}, \ion{H}{$\beta$} and
\ion{H}{$\gamma$} show P Cyg behaviour which indicate ongoing
post-AGB mass-loss. Based on ASAS and ASAS-SN data, we have
detected rapid photometric variability in Hen~3-1013 with an
amplitude up to 0.2 mag in the $V$ band. The star's low-resolution
spectrum underwent no significant changes from 1994 to 2012. Based
on archival data, we have traced the photometric history of the
star over more than 100 years. No significant changes in the star
brightness have been found.

\end{abstract}


\section{INTRODUCTION}\label{sec:intro}

The southern star CPD-64\degr2939 was identified by \citet{hen76}
as an \ion{H}{$\alpha$} emission line object and was named
Hen~3-1013. \citet{partpot89} found it to be an IRAS source with
far infrared (IR) colours similar to planetary nebulae (PNe) and
suggested that it is in the post-AGB stage of evolution.
\citet{loup90} detected CO emission in this object and derived an
expansion velocity of 15 \kms. From low resolution optical spectra
\citet{pvd00} find it to be a B3 supergiant with H$\beta$ in
emission and H$\gamma$ filled in. In 2003 \citet{gp03} have
estimated the interstellar and circumstellar absorption of Hen
3-1013 on the low resolution \emph{IUE} spectrum  from 1150 to
3200 \AA. Combining the optical, near and far-IR (\emph{ISO},
\emph{IRAS}) data of IRAS~14331-6435 \citet{gp04} have
reconstructed its spectral energy distribution (SED) and estimated
the dust temperatures, mass loss rates, angular radii of the inner
boundary of the dust envelopes and the distances to the star. In
\emph{ISO} spectrum  these authors have found the amorphous and
crystalline silicate features indicating oxygen-rich circumstellar
dust shell of the star. The identical features  were observed in
oxygen-rich post-AGB stars IRAS~18062+2410 (V886~Her) and IRAS
22023+5249 (LSIII +52~24).

\citet{garcia02} observed He~3-1013 in 2~$\mu m$ H$_{2}$ survey,
have found the emission lines of H$_{2}$ molecule and measured the
flux ratio of H$_{2}$ $[v=1\rightarrow 0~
\mathrm{S(1)}/2\rightarrow1~\mathrm{S(1)}]=8.4$ what showed the
collisional excitation H$_{2}$ in He~3-1013.

\citet{mello12} have determined the chemical abundances of
He~3-1013 on high resolution spectra obtained on the 2.2-m
telescope at La Silla with the FEROS spectrograph. They received
the stellar parameters by using of non-LTE model atmospheres
BSTAR2006 \citep{lanz07} and estimated the core mass and zero-age
mass of the star from evolutionary tracks of post-AGB stars
according to \citet{blocker95}. The core mass turned out to be
$0.70\pm0.20$M$_{\odot}$ and the metallicity Z(CNO)=0.016.

The masses of stars on the post-AGB evolution stage are
in the range 0.5-0.8 M$_{\odot}$. If Hen~3-1013 indeed
is related to more massive post-AGB stars
($\sim$ 0.7 M$_{\odot}$) one would expect possible rapid evolution.
In this paper, we traced the photometric history of Hen~3-1013 over more
than 100 years and analysed the spectroscopic data over the
last 20 years to test this assumption.

The optical spectrum of Hen~3-1013 has so far not been studied
in sufficient detail. In this paper, we report the results of
the high-resolution spectral observations carried out with
the aim to study the peculiarities of the spectrum and
the details of the velocity field in the atmosphere and envelope
of the star.

In Section~\ref{sec2} we briefly describe the spectral
observations of Hen~3-1013 and the data reduction process. A
detailed analysis of the high-resolution spectrum is presented in
Section~\ref{sec3} while the analysis of the low-resolution
spectrum is shown in Section~\ref{sec4}. A photometric history of
Hen~3-1013 is presented in Section~\ref{sec5}. We discuss our findings
and conclude in Section~\ref{sec6}.

\section{OBSERVATIONS AND DATA REDUCTION}\label{sec2}

\subsection{High-resolution spectrum}

The  spectrum of Hen~3-1013 was obtained on April 14, 2006 with
the FEROS spectrograph \citep{kauf99} and MPI/ESO 2.2-m telescope
(Proposal ID. 77.D-0478A, PI: M.Parthasarathy). The resolving
power is $R\sim48\,000$ and the wavelength coverage is from 3600
to 9200 \AA. During a 45 min exposure we achieved a
signal-to-noise ratio of about 100 at the wavelength of about 5400
\AA. The reduction of the FEROS spectrum was performed by the
on-line  software,  including  flat-field correction,  background
subtraction, removal of cosmic rays, wavelength calibration,
barycentric velocity correction, and continuum normalisation. More
details of instrument set up, observations and data reduction can
be found in \citep{otsuka17}.

 \subsection{Low resolution spectra}

 Our low resolution spectroscopic observations  were carried out in
 11 May 2012 at the 1.9-m telescope of the South African
 Astronomical Observatory (SAAO) with a long-slit spectrograph at
 the Cassegrain focus. The slit was $\sim3\arcmin$ in length and
 $1.\arcsec5$ in width; the scale along the slit was
 $0.\arcsec7$/pixel. The detector was an SITe 266$\times$1798-pixel
 CCD array. A 300 lines/mm grism was used in the spectral range
 3500-7200 \AA. The actual spectral resolution was FWHM=4.5 \AA.
 Spectra of a Cu-Ar-filled lamp were taken to calibrate the
 wavelengths after each observation. Bias and flat-field images
 were also obtained for each night of observations to perform the
 standard reduction of two-dimensional spectra.

 In addition, we used the spectroscopic data for Hen~3-1013 from
 the appendix to the "Spectroscopic atlas of post-AGB stars and
 planetary nebulae"\ by \citet{suarez06}. The observations were
 carried out in Chile with a 1.5-m telescope at the La Silla
 Observatory of the European Southern Observatory (ESO) using the
 Boller-Chivens spectrograph. The formal resolution was 3.74 \AA\/
 pixel. The spectrum was taken in the period March 13-17, 1994, in
 the spectral range 3285-10~980 \AA.

 \section{ANALYSIS OF THE HIGH RESOLUTION SPECTRUM}\label{sec3}

 \subsection{Description of the spectrum}

 The optical spectrum of Hen~3-1013 displays stellar absorption
 lines and nebular emission features. The complete
 continuum-normalised and smoothed spectrum of Hen~3-1013 in the
 spectral ranges 3700--7200, 7340--7540, 7720--7920 and 8330--8530
 \AA\ is presented in Appendix A (Fig.~\ref{fig:sp}). We used the
 adjacent averaging by 10 points to smooth the spectrum.

 The search and identification of the lines in spectrum of
 Hen~3-1013 was carried out by using of spectral orders of echelle
 normalised to stellar continuum starting from 3800 to 9000 \AA.
 The standard wavelengths of the lines were taken in the main from
 the tables by \citet{moore45}, the National Institute of Standards
 and Technology (NIST) Atomic Spectra
 Database\footnote{https://www.nist.gov/pml/atomic-spectra-database}
 and the \ion{Fe}{II}-data by \citet{nave12}.

 \subsubsection{Nebular emission lines}\label{emlines1}

 \begin{table}
 \begin{center}
  \caption{Emission lines in Hen~3-1013.}
 \label{emlines}

 \begin{tabular}{ccccc}

 \hline

   $\lambda_\text{obs.}$& $\lambda_\text{lab.}$&Identification&
 $W_\lambda$& $V_\text{r}$\\
 (\AA) & (\AA) &&(\AA)& ({\kms}) \\

  \hline

 3853.12&3853.66&\ion{Si}{II}(1)& 0.033& -42.82\\
 3855.42&3856.00&\ion{Si}{II}(1)& 0.044& -46.68\\
 4200.33&4200.89&\ion{Si}{II}&0.024&-39.99\\
 4243.47&4243.97&[\ion{Fe}{II}](7F)&0.026&-35.32\\
 4286.93&4287.39&[\ion{Fe}{II}](7F)&0.076 &-30.79\\
 4358.79&4359.34&  [\ion{Fe}{II}](7F)&0.040&-37.85\\
 4390.01&4390.58&\ion{Mg}{II}(10)&0.022&-38.95\\
 4413.20&4413.78&[\ion{Fe}{II}](7F)&0.032&-39.39\\
 4433.46&4433.99&\ion{Mg}{II}(9)&0.019&-35.86\\
 4457.41&4457.95&[\ion{Fe}{II}](6F)&0.020&-36.31\\
 4583.23&4583.83&\ion{Fe}{II}(38)&0.025&-39.27\\
 4813.97&4814.55&[\ion{Fe}{II}](20F)&0.041&-36.14\\
 4889.16&4889.63&[\ion{Fe}{II}](3F), 4F~bl&0.022&-28.84\\
 4904.86&4905.35&[\ion{Fe}{II}](20F)&0.017&-29.97\\
 5040.35&5041.03&\ion{Si}{II}(5)&0.030&-40.47\\
 5055.27&5055.98&\ion{Si}{II}(5)&0.043&-42.13\\
 5158.13&5158.81&[\ion{Fe}{II}](19F)&0.051&-39.54\\
 5184.97&5185.56&\ion{Si}{II}(5)&0.031&-34.13\\
 5197.29&5197.90&[\ion{N}{I}], \ion{Fe}{II}(49)&0.041&-35.21\\
 5261.02&5261.61&[\ion{Fe}{II}](19F)&0.037&-33.64\\
 5275.26&5275.82&[\ion{Ni}{II}], \ion{Fe}{II}(49)&0.021&-31.84\\
 5465.92&5466.45&\ion{Si}{II}&0.071&-29.09\\
 5956.84&5957.57&\ion{Si}{II}(4)&0.041&-36.76\\
 5978.27&5978.93&\ion{Si}{II}(4)&0.107&-33.12\\
 6231.06&6231.75&\ion{Al}{II}(10)&0.057&-33.22\\
 6238.86&6239.60&\ion{Si}{II}(4)&0.082&-35.58\\
 6242.67&6243.36&\ion{Al}{II}(10)&0.074&-33.16\\
 6248.12&6248.91&\ion{Fe}{II}(74)&0.022&-37.93\\
 6317.19&6317.98&\ion{Fe}{II}(J)&0.070&-37.51\\
 6442.14&6442.97&\ion{Fe}{II}(J)&0.040&-38.62\\
 6666.04&6666.89&[\ion{Ni}{II}](2F)&0.035&-38.25\\
 6822.69&6823.48&\ion{Al}{II}(9)&0.015&-34.73\\
 6828.99&6829.83&\ion{Si}{II}&0.021&-36.90\\
 6836.34&6837.14&\ion{Al}{II}(9)&0.038&-35.10\\
 6851.01&6851.89&\ion{Fe}{II}(J)&0.017&-38.50\\
 7041.15&7042.06&\ion{Al}{II}(3)&0.097&-38.77\\
 7055.67&7056.60&\ion{Al}{II}(3)&0.054&-39.54\\
 7377.06&7377.83&[\ion{Ni}{II}](2F)&0.249&-31.31\\
 7410.75&7411.61&[\ion{Ni}{II}](2F)&0.053&-34.81\\
 7494.75&7495.63&\ion{Fe}{II}(J)&0.059&-35.22\\
 7505.70&7506.54&\ion{Fe}{II}(J)&0.064&-33.57\\
 7512.30&7513.18&\ion{Fe}{II}(J)&0.154&-35.14\\
 7730.85&7731.67&\ion{Fe}{II}(J)&0.079&-31.80\\
 7847.82&7848.80&\ion{Si}{II}&0.086&-37.44\\
 7848.78&7849.72&\ion{Si}{II}, \ion{Fe}{II}&0.121&-35.93\\
 7876.02&7877.05&\ion{Mg}{II}(8)&0.268&-39.23\\
 7895.31&7896.37&\ion{Mg}{II}(8)&0.409&-40.27\\
 7999.08&8000.07&[\ion{Cr}{II}](1F)&0.140&-37.12\\
 8124.31&8125.30&[\ion{Cr}{II}](1F)&0.087&-36.55\\
 8213.11&8213.99&\ion{Mg}{II}(7)&0.202&-32.14\\
 8233.63&8234.64&\ion{Mg}{II}(7)&0.265&-36.80\\
 8286.72&8287.59&\ion{Fe}{II}(J)&0.200&-31.49\\
 8307.48&8308.51&[\ion{Cr}{II}]&0.070&-37.19\\
 8445.44&8446.25&\ion{O}{I}&1.039&-28.77\\
 8450.06&8451.01&\ion{Fe}{II}(J)&0.144&-33.72\\
 8489.13&8490.10&\ion{Fe}{II}(J)&0.157&-34.28\\
 8616:&      8616.96&[\ion{Fe}{II}](13F)&--&--\\
 8628.21&8629.24&\ion{N}{I}(8)&0.081&-35.81\\
 8635.53&8636.58&\ion{Fe}{II}(J)&0.075&-36.47\\
 8639.64&8640.70&\ion{Al}{II}(4)&0.062&-36.80\\
 8925.81&8926.90&\ion{Fe}{II}, \ion{Al}{II}&0.248&-36.63\\

  \hline
 \end{tabular}
 \end{center}
 \end{table}

 The list of emission lines in Hen 3-1013 is given in
 Table~\ref{emlines}. It includes the measured and laboratory
 wavelength (in the air), the equivalent width ($W_\lambda$), the
 radial velocity ($V_\text{r}$) with respect to Sun, the name of the
 element and the multiplet number to which the measured line
 belongs. The hydrogen lines in the table are missed owing to they
 have the complex multicomponent profiles and  will be discussed
 separately.

 The permitted  emission lines, in addition to hydrogen, belong  to
 the ions of \ion{Si}{II} (from 1, 4, 5 multiplets), \ion{Mg}{II}
 (7, 8, 9, 10), \ion{Fe}{II} (38, 42, 49), \ion{Al}{II} (3, 4, 9,
 10),  and also to the nonionised atoms of \ion{O}{I} (3),
 \ion{N}{I} (8).

 The forbidden emission lines are represented mainly by the lines
 of [\ion{Fe}{II}]  from  multiplets 6F, 7F, 19F, 20F, 21F and
 others, [\ion{Ni}{II}] - by three lines in the red and
 near-infrared of 2F multiplet and in visible - from 14F multiplet.
 In the range of 8000-8300 \AA\ the lines of [\ion{Cr}{II}] of  1F
 multiplet are observed. The forbidden line of [\ion{N}{I}] $\lambda$5198
 are presented. In the red and the near infrared there are
 seen the numerous emissions \ion{Fe}{II} from  the
 intercombination transitions (designated as J) whose excitation
 usually put down to the fluorescence due to the \ion{Ly}{$\alpha$}
 or to the ultraviolet continuum of the star.

 We have estimated the intensities of two forbidden lines of
 [\ion{Ni}{II}] from  2F multiplet  in the near-IR of
 He~3-1013. Their flux ratio was found to be
 $I(7411)/I(7378)=0.20\pm 0.01$.  In the pure collisional
 excitation this value is about 0.1 \citep{baut96} if the electron
 density $N_e <10^6$ cm$^{-3}$ (by $T_{e}\sim 10000$ K).
 \citet{lucy95} had studied the cause of anomalously strong
 emission the lines at $\lambda\lambda$6667, 7378 and 7412 of 2F multiplet
 [\ion{Ni}{II}]  observable in a variety gaseous nebulae and
 suggested the photon pumping of these by background UV continuum.
 \citet{baut96}  have computed the line ratio $I(7411)/I(7378)$ in
 the three-level model of [\ion{Ni}{II}] depending  on the electron
 density of ambient gas and  have  shown that if $N_{e}\geq10^7$
 cm$^{-3}$ the photon pumping missed. The gaseous envelope  of
 He~3-1013 very  probably may be the partially ionized zone (PIZ)
 and may have the high enough electron density ($N_{e}=10^6-10^7$ cm$^{-3}$). The
 reasonable difference of radial velocities found on emission and
 absorption lines  in star spectrum allows us to assume this. The
 comparison of the observed line ratio (without interstellar
 reddening)  and the theoretical one (\citet{baut96}, Fig.3)
 testified to the pure collisional excitation of [\ion{Ni}{II}]
 lines in the gaseous shell of Hen~3-1013 with $N_{e}$ near 10$^7$ cm$^{-3}$.

 The mean radial velocities of the emission lines for various
 elements are shown in Table~\ref{meanvr}. In this table $N$
 designates the number of lines used to take the mean $V_\text{r}$.
 Particularly we note that the mean $V_\text{r}$ of the emission lines in
 the spectrum of Hen~3-1013 are practically the same for all
 revealed chemical elements (except for hydrogen) and that the
 difference of $V_\text{r}$ between the permitted and forbidden lines of
 the same element is absent.

 On the whole the average heliocentric  radial velocity of the
 emission envelope in Hen 3-1013 turned out to be
 $V_\text{r}=-36.0\pm0.4$ {\kms}.

 We measured nebular expansion velocity ($V_\text{exp}$) using the
 following relation: $V_{\text{exp}}=1/2(V_\text{FWHM}^2-V_\text{instr}^2)^{1/2}$, where
 $V_\text{FWHM}$ is the velocity corresponding to the
 full width at half maximum (FWHM) and $V_\text{instr}$ (6 \kms) is the
 instrumental broadenings. For the [\ion{Ni}{II}]
 $\lambda$7378 emission line $V_\text{FWHM}$ equal to 25.5 \kms.
 The expansion velocity for Hen~3-1013 from this line is 12.4 \kms.
 From CO observations \citet{loup90} estimated expansion velocity
 of 15 \kms.

 \begin{table}
 \begin{center}
  \caption{The mean heliocentric radial velocity of emission
  lines  of selected ions in  Hen 3-1013.}
 \label{meanvr}

 \begin{tabular}{ccc}

 \hline
 ion& $V_\text{r}$ (\kms)&$N$\\
 \hline
 \ion{Fe}{II}&$-35.51\pm 0.68$&14\\
 \ion{Si}{II}&$-37.90\pm 1.39$&12\\
 \ion{Mg}{II}&$-37.18\pm 1.21$&6\\
 \ion{Al}{II}&$-35.88\pm 0.96$&7\\
 $[\ion{Ni}{II}]$&$-34.77\pm 2.00$&3\\
 $[\ion{Fe}{II}]$&$-35.42\pm 1.15$&9\\
 $[\ion{Cr}{II}]$&$-36.93\pm 0.20$&3\\

  \hline
 \end{tabular}
 \end{center}

 \end{table}

\subsubsection{The hydrogen lines}

\begin{figure}
 \includegraphics[width=\columnwidth]{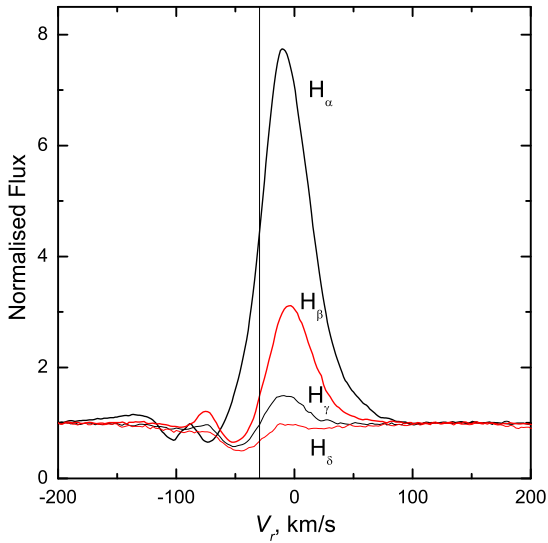}
 \caption{Balmer lines in the spectra of Hen~3-1013. The velocities
 are in the heliocentric frame. The vertical line marks the radial
 velocity determined from absorption lines.}
 \label{balm}
\end{figure}

Our material show Balmer lines from \ion{H}{}14 to \ion{H}{}7 in
absorption (see Table.~\ref{abslines}). The \ion{H}{$\alpha$},
\ion{H}{$\beta$} and \ion{H}{$\gamma$} lines are dominated by
emission component. The \ion{H}{$\delta$} line is also partly
contaminated by the emission component. As to emission lines of
hydrogen so the profiles of \ion{H}{$\alpha$}-\ion{H}{$\delta$}
lines are shown at Fig.~\ref{balm} where the velocities are in the
heliocentric frame. The strong \ion{H}{$\alpha$} emission with
broad wings has $V_\text{r}=-7.8$ \kms and the red wing of this
line extend more than 100 km/s. In the blue wing of
\ion{H}{$\alpha$} there are seen two absorption components on
$V_\text{r}(1)=-70$ \kms and $V_\text{r}(2)=-100$ \kms.
Uncertainty in the $V_\text{r}$ is about 1 \kms. The other
\ion{H}{I} emissions have: $V_\text{r}$(\ion{H}{$\beta$})=--7.4
\kms, $V_\text{r}$(\ion{H}{$\gamma$})=--9.0 \kms,
$V_\text{r}$(\ion{H}{$\delta$})=--6.0 \kms. The equivalent widths
of the emission components of \ion{H}{$\alpha$}, \ion{H}{$\beta$}
are equal to 7.63 and 1.65 \AA\ respectively.

We compared the \ion{H}{$\alpha$} profile on our spectrum and on
the spectrum obtained on the same telescope with the same
spectrograph in September 2007 or May 2008 \citep{mello12}. It
should be noted that our \ion{H}{$\alpha$} profile is different
from that previously reported by \citet{mello12}. In particular,
on our spectrum, the wings are narrower and there is a weak
emission component at wavelength 6560.9 \AA\, which is not in the
spectrum from \citet{mello12}.

Pashen lines on our spectrum are presented by high members from
P16 and more. The line profile of each of them consist of
the absorption component and more weak emission disposed
on the blue wing of absorption. The mean velocity of Pashen
absorptions is near Balmer emission and is about --8 \kms.

\subsubsection{Photospheric absorption lines}

All of the identified absorption lines together with their radial
velocity and equivalent weight are summarized in
Table~\ref{abslines}. In the column 1 the measured wavelength of
the line is given, the columns 2-3 - the standard wavelength of
identified ion (of the primary component if the blend) with
multiplet number, the columns 4-5 - the possible member of the
line if  in the blend, the column 6 - the equivalent width of
primary absorption line, the column 7 - the radial velocity of the
single component or the primary component in the blend.

Absorption lines of neutral species including \ion{H}{i},
\ion{He}{i}, \ion{C}{I}, \ion{N}{I}, \ion{O}{I}  and \ion{Ne}{I}
were identified. Singly-ionized species including \ion{C}{II},
\ion{N}{II}, \ion{O}{II}, \ion{Si}{II}, \ion{S}{II}, \ion{Ar}{II},
\ion{Fe}{II}, \ion{Mn}{II}, \ion{Cr}{II}, \ion{V}{II},
\ion{Ti}{II}, \ion{Co}{II}, \ion{Ni}{II}, \ion{Al}{II},
\ion{Cu}{II} and \ion{Cs}{II} were detected. Higher ionization is
seen in \ion{Si}{III}, \ion{S}{III}, \ion{Fe}{III} and \ion{Si}{IV}.

The mean radial velocities of the absorption lines for various
elements are shown in the Table~\ref{meanvrabs}. The comparison of
mean values $V_\text{r}$ absorption lines of the various elements may
point out on some stratification of the chemical element's
velocity in the atmosphere of Hen 3-1013: for example, $V_\text{r}$ of
\ion{S}{II} and \ion{Si}{II} absorptions distinguish from average
$V_\text{r}$ of other absorptions so far as 5 \kms.

\begin{table}
\begin{center}
 \caption{The mean heliocentric radial velocity of absorption
 lines  of selected ions in  Hen 3-1013.}
\label{meanvrabs}

\begin{tabular}{ccc}

\hline
ion& $V_\text{r}$ (\kms)&$N$\\
\hline

\ion{He}{i}&$-23.66\pm 1.29$&22\\
\ion{O}{I}&$-31.61\pm 1.37$&7\\
\ion{O}{II}&$-31.35\pm 0.62$&48\\
\ion{N}{II}&$-32.17\pm 0.63$&34\\
\ion{C}{II}&$-29.21\pm 1.44$&13\\
\ion{S}{II}&$-25.92\pm 0.62$&30\\
\ion{Fe}{II}&$-32.49\pm 1.34$&13\\
\ion{Si}{II}&$-24.62\pm 1.28$&6\\
\ion{Ar}{II}&$-33.28\pm 1.31$&7\\
\ion{S}{III}&$-27.16\pm 1.80$&7\\
\ion{Fe}{III}&$-29.27\pm 2.54$&7\\
\ion{Si}{III}&$-31.24\pm 1.74$&8\\
\ion{Si}{IV}&$-26.31\pm 0.12$&2\\

 \hline
\end{tabular}
\end{center}

\end{table}

Radial velocities measured from non-blendend lines are
$-29.6\pm0.4$ {\kms}, which is in good agreement with the previous
estimate of $-29.8\pm2.1$ {\kms} \citep{mello12}.

\subsubsection{Interstellar features and colour excess}

\begin{figure}
 \includegraphics[width=\columnwidth]{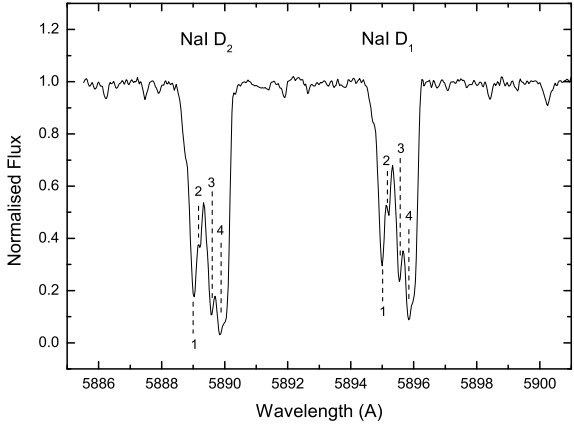}
 \caption{\ion{Na}{I} D$_{2}$ and \ion{Na}{I} D$_{1}$ lines in
  the spectra of Hen~3-1013. The various absorption
  components have been labelled.}
 \label{fig:dna}
\end{figure}

The \ion{Na}{I} double resonance D-lines in the high-resolution
spectrum of Hen~3-1013 show a complex profiles. Four absorption
components were identified in the \ion{Na}{i} D2 and \ion{Na}{i}
D1 lines (see Fig.~\ref{fig:dna} and Table~\ref{NaI}). The
velocities of the component 2: $V_\text{r}=-37.0\pm0.8$\kms are
comparable with the mean heliocentric radial velocity of the
emission lines in the star $V_\text{r}=-36.0\pm0.4$\kms,
suggesting that this component arises in an extended envelope
around the central star. We may infer that 1, 3 and 4 components
in the velocity interval from --5 to --47\kms~originate in the
interstellar medium.

\begin{table}
\begin{center}
 \caption{Absorption components of \ion{Na}{I} D$_{2}$ (5889.95 \AA) and
 \ion{Na}{I} D$_{1}$ (5895.92 \AA) lines in the spectrum of Hen~3-1013.
 $V_\text{r}$ are the respective heliocentric radial velocities.}
 {\small

\label{NaI}
\begin{tabular}{ccccc}
  \hline
  component&\multicolumn{2}{c} {\ion{Na}{I} D$_{2}$} & \multicolumn{2}{c} {\ion{Na}{I} D$_{1}$}\\
  & $\lambda_\text{obs.}$&$V_\text{r}$&$\lambda_\text{obs.}$&$V_\text{r}$\\
  &(\AA)& (\kms)&(\AA)& (\kms)\\
  \hline

  1&5889.03&--47.01&5895.00&--46.96\\
  2&5889.21&--37.84&5895.21&--36.28\\
  3&5889.57&--19.51&5895.54&--19.49\\
  4&5889.84&--5.76&5895.84&--4.22\\

  \hline
\end{tabular}
}
\end{center}
\end{table}

In the spectrum Hen~3-1013 we identified the most famous Diffuse
Interstellar Bands (DIBs) \citep{Hobbs08} (see Table~\ref{DIB}).
We estimated an interstellar $E(B-V)=0.43$ mag from the measured
$W_{\lambda}=0.22$ \AA\ for the $\lambda$5780 using the
correlation $W_{\lambda}$ with $E(B-V)$ obtained by
\citet{fried11}.

The interstellar extinction estimate of Hen~3-1013
from the 2200 \AA\ feature in the UV: $E(B-V)=0.44$ mag \citep{gp03}
is in good agreement with the colour excess $E(B-V)$ was determined
by us from DIB $\lambda$5780. The total colour excess of Hen 3-1013 turned
out as $E(B-V)=0.71$ mag \citep{gp03} hence the star have observable
circumstellar extinction.

\begin{table}
\begin{center}
 \caption{DIBs in Hen~3-1013}
 {\small

\label{DIB}
\begin{tabular}{crcr}
  \hline

$\lambda_\text{obs.}$ (\AA) & $\lambda_\text{lab.}$ (\AA) &FWHM (\AA)&
$W_{\lambda}$ (\AA)\\

  \hline
4430.40& 4428.19& 1.83 & -0.12\\
5485.20& 5487.69& 0.92 & -0.02\\
5704.70& 5705.08& 2.81 & -0.12\\
5780.37& 5780.48& 1.98 & -0.22\\
5796.87& 5797.06& 0.87 & -0.04\\
6195.80& 6195.98& 0.84 & -0.03\\
6203.81& 6203.05& 2.14 & -0.09\\
6269.67& 6269.85& 1.25 & -0.04\\
6283.65& 6283.84& 5.50 & -0.82\\
6379.05& 6379.32& 0.87 & -0.06\\
6613.48& 6613.62& 0.97 & -0.08\\

 \hline
\end{tabular}
}
\end{center}
\end{table}

\subsection{Parameters of the stellar atmosphere}

To determine the main model parameters of the atmosphere, such as
the effective temperature and gravity, we used non-LTE,
plane-parallel, hydrostatic model atmosphere BSTAR2006 grid
\citep{lanz07} in the TLUSTY package \citep{hubeny95}. We first
estimate a surface gravity for the sample star using Balmer
\ion{H}7 $\lambda$3835. For this exercise, we adopt an initial
guess of $T_{\text{eff}}=19~000$~K, then searched for a $\log g$
value that produce observed profile of the \ion{H}7 line. As a
result of this exercise, we have obtained $\log g=2.3$. We then,
estimated $T_{\text{eff}}$ based on \ion{Si}{III}/\ion{Si}{IV}
ionization balance. This exercise yields $T_{\text{eff}}=17~750$
K. We have used \ion{Si}{} and \ion{O}{} absorption lines to
estimate the micro-turbulent  and rotational velocity. This
results in $\xi_{t}=24$ \kms and $v \sin i=5$\kms.

The initial estimations of $T_{\text{eff}}$, $\log g$ and
$\xi_{t}$ have been iterated until a consistent set of parameters
are obtained. We have finally obtained the values:
$T_{\text{eff}}=18~250\pm500$ K, $\log g=2.3\pm0.05$ dex,
$\xi_{t}=32\pm3$ \kms and $v \sin i=5\pm1$\kms.

\citet{mello12} obtained $T_{\text{eff}}=16~200\pm300 K$, $\log
g=2.04$, $\xi_{t}=17$ \kms and $v \sin i=38\pm5$\kms. The
rotational velocity obtained by them is very high and $\xi_{t}$
obtained by them is low compared to the value that we obtained.
This may the reason for differences in $T_{\text{eff}}$ values. A
careful reanalysis of all the absorption line profiles is needed,
which is beyond the scope and aim of the present paper. In this
range of temperatures of B stars, the effects of $T_{\text{eff}}$
and surface gravity on the wings of Hydrogen lines are degenerate:
roughly similar \ion{H}{} theoretical profiles can be computed
with pairs ($T_{\text{eff}}$; $\log g$) and
($T_{\text{eff}}$--1~000 K; $\log g$--0.1), for example. This
degeneracy, in principle, may explain the differences between the
results obtained in the present paper and those published by
\citet{mello12}.

\citet{gp03} from an analysis of the UV (\emph{IUE}) low
resolution spectra derived $T_{\text{eff}}=16200$ K and $\log
g=2.6$ which are in reasonable agreement with those derived by
\citet{mello12}. The $T_\text{eff}$ and $\log g$ values derived
from the UV flux distribution are sensitive to interstellar
reddening and circumstellar reddening values.

\section{LOW-RESOLUTION SPECTROSCOPY}\label{sec4}

\begin{figure*}
 \includegraphics[scale=1.2]{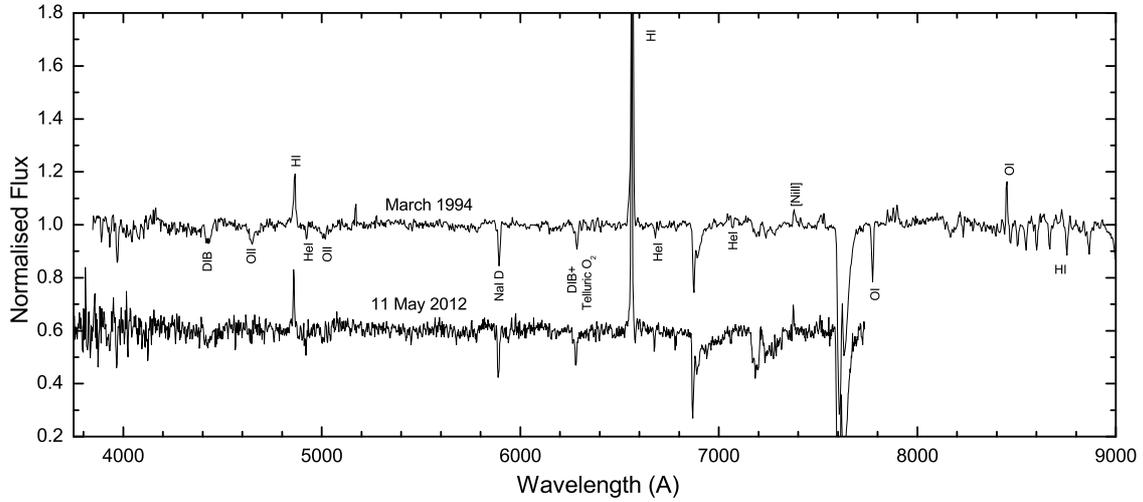}
 \caption{The low-resolution optical spectra of Hen 3-1013
 from \citet{suarez06} (top) and obtained in SAAO (bottom)
 normalised to the continuum. The principal lines and most
 prominent DIBs are indicated. For clarity, the spectra are
 offset by 0.4 continuum flux unit.}
 \label{fig:sp_low}
\end{figure*}

Fig.~\ref{fig:sp_low} presents the low-resolution spectra of Hen
3-1013 obtained by \citet{suarez06} in 1994 and by us in 2012. In
contrast to the high-resolution spectrum, these spectra show only
the strongest lines, namely, the emission lines \ion{H}{$\alpha$},
\ion{H}{$\beta$}, \ion{O}{I} $\lambda$8446, [\ion{Ni}{II}]
$\lambda$7378 and the absorption lines  \ion{O}{I} $\lambda$7774,
the \ion{Na}{I} D interstellar resonance lines, DIBs at
$\lambda$4430 and $\lambda$6282. One of our interests is to check
whether the spectrum has changed from 1994 to 2012.

We measured the equivalent widths of the strongest emission and
absorption lines in the low-resolution spectrum of Hen~3-1013 and
provide them in Table~\ref{splow}. The line equivalent widths in
the spectrograms taken at different dates agree well between
themselves. Thus, the star's spectrum underwent no significant
changes from 1994 to 2012. The emissions of [\ion{Ni}{II}]
$\lambda$7378 and \ion{H}{$\alpha$} on low-resolution spectra are
stronger than in high-resolution spectra. These differences may be
associated to a greater extent with the difference in spectral
resolution.

\begin{table}
\begin{center}
\caption{Equivalent widths of several lines in the low-resolution
spectrum of Hen 3-1013. } \label{splow}
{\small
\begin{tabular}{cccc} \hline $\lambda$ (\AA) & Ion
&\multicolumn{2}{c} {$W_{\lambda}\pm \sigma_{W_{\lambda}}$ (\AA)} \\
&&1994-03 & 2012-05-11 \\ \hline

4861 & \ion{H}{$\beta$} & 2.1$\pm$0.2 & 1.8$\pm$0.2\\
6563 & \ion{H}{$\alpha$}& 17.1$\pm$0.6 & 15.0$\pm$0.4\\
6678 & \ion{He}{I} &-0.40$\pm$0.08&-0.40$\pm$0.10\\
7378 & [\ion{Ni}{II}] & 0.76$\pm$0.08&0.60$\pm$0.10\\

 \hline
\end{tabular}
}
\end{center}

\end{table}

\section{PHOTOMETRIC VARIABILITY AND HISTORY OF Hen~3-1013}\label{sec5}

Hen~3-1013 is present in the Cape Photographic Durchmusterung
(CpD) \citep{gill00} as CpD-64\degr2939 with $m_{pg}=9.7$ mag, which
allowed the photometric history of the star to be traced over more
than 100 years. However, photographic magnitudes $m_{pg}$ from the
CpD catalogue should be analysed before they are compared with the
present-day photometric data.

\begin{figure}
 \includegraphics[width=\columnwidth]{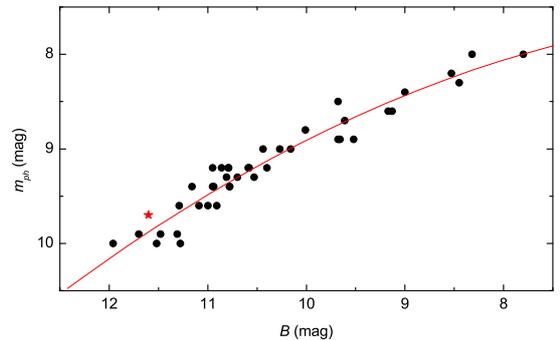}
 \caption{Relationship between the $B$ magnitudes from SIMBAD and
 the photographic magnitudes from the CpD for Hen~3-1013 (asterisk) and
 the stars from its neighbourhood (dots). The dashed lines indicate
 parabola fits to the data for the stars from the neighborhoods.}
 \label{fig:hist}
\end{figure}

We selected 41 stars contained in the CpD from a $1^{\circ} \times
1^{\circ}$ neighbourhood of Hen~3-1013 and derived the
relationship between the $B$ magnitudes from SIMBAD and the
photographic magnitudes $m_{ph}$ from the CpD of Hen~3-1013 and
stars from its neighbourhood (Fig.~\ref{fig:hist}). The
relationships between $B$ and $m_{pg}$ of the stars from the
neighbourhood obey the quadratic polynomial law, according to
which $m_{pg}$ of Hen~3-1013 from the CpD is transformed into the
$B$ magnitude $\sim11.4$ mag.

During 1973 and 1974 2 $UBV$ measurements of the star were
obtained by \citet{kn77}. Hen~3-1013 was also observed by
\citet{kozok85} from 1979 (2 nights) to 1980 (3 night). According
to the $UBV$ data from \citet{kozok85}, the brightness Hen 3-1013
changed by 0.1 mag, with the error of a single observation being
0.010-0.015 mag.

Hen~3-1013 fell within the field of view of the All Sky Automated
Survey (ASAS) \citep{pojm02}. The observations in the ASAS-3
system have been carried out in 2001-2009 at the Las Campanas
(Chile) telescopes in an automatic mode in a photometric $V$ band
close to Johnson's standard $V$. To analyse the ASAS-3 data, we
used the measurements made with aperture 1 (15$\arcsec$) and
marked in the ASAS-3 database by symbol A (good quality). The mean
accuracy of the measurements was 0.04 mag.

All Sky Automated Survey for Super-Novae (ASAS-SN) is an all-sky
survey that monitors the entire sky for transients every night to
a depth of $V$-band of $\sim$17 mag \citep{shap14, koch17}. The
field containing Hen~3-1013 was observed a total of 505 times from
2016 March 10 to 2018 May 7 by \textsf{bh} camera in the station,
located at the Cerro Tololo International Observatory (CTIO,
Chile). The mean photometric uncertainty for these choices is
0.008 mag.

\begin{figure}
 \includegraphics[width=\columnwidth]{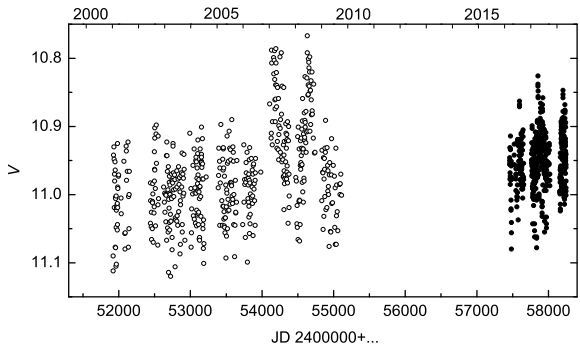}
 \caption{ASAS (the open circles) and ASAS-SN (the filled circles)
  $V$ band light curve of Hen~3-1013 from 2001 to 2018}.
 \label{fig:asas}
\end{figure}

Fig.~\ref{fig:asas} shows the light curves derived from ASAS-3 and
ASAS-SN data. The pattern of variability for Hen~3-1013 is similar
in both characteristic time scales of brightness variations and
oscillation amplitudes. The mean brightness in $V$-band from ASAS
and ASAS-SN are listed in Table~\ref{tab:hist}.

The star displays a brightness variation with an amplitude
(peak-to-peak) of up  to 0.2 mag in $V$-band. Light variability
may be considered real since 3$\sigma$-threshold is exceeded
several times.

The ASAS light curve from 2001 to 2006 and in 2009 (see
Fig.~\ref{fig:asas}) shows a short time-scale irregular
variability with  an amplitudes of 0.1-0.2 mag and roughly equal
seasonal means ($V\sim11.0$ mag).

ASAS data in 2007 display a trend of decreasing brightness. In
2008, the star's brightness systematically increased. Superimposed
on this trends is a clear fast variation with amplitude up  to
0.15 mag. The mean brightness in 2007-2008 is 0.1 mag higher, when
compared to the earlier and later ASAS measurements (see
Table~\ref{tab:hist}).

According to higher precision and better time-resolution ASAS-SN
data in 2016-2018, the star brightness changes more regularly with
characteristic time from 6 to 15 days in different seasons. The
peak-to-peak range varies from 0.05 to 0.17 mag in $V$. We failed
to determine the period both in the total dataset and in the
subsets of single seasons. The variability of Hen 3-1013 is very
similar to that of hot post-AGB star IRAS~19336-0400
\citep{arkh12}.

\begin{table}
 \caption{Summary of photometric observations of Hen~3-1013.}
 {\small
 \label{tab:hist}
 \begin{tabular}{lcccc}
  \hline
  Date& $U$ (mag)& $B$ (mag)& $V$ (mag)& source\\
  \hline
  before 1900& --&$\sim$11.4 &--&(1)\\
  1973-1974&11.128&11.485&10.889&(2)\\
  1979-1980&$11.112$&$11.479$&$10.899$&(3)\\
  2001-2006&--&--&$11.00\pm0.05$&ASAS-3\\
  2007-2008&--&--&$10.91\pm0.07$&ASAS-3\\
  2009&--&--&$10.99\pm0.05$&ASAS-3\\
  2016-2018&--&--&$10.95\pm0.04$&ASAS-SN\\
  \hline
 \end{tabular}
 }(1) -- \citet{gill00}; (2) -- \citet{kn77}; (3) --
 \citet{kozok85}.
\end{table}

In Table~\ref{tab:hist}, together with ASAS and ASAS-SN data, the
the transformed value from CpD, the average $UBV$ magnitudes from
\citet{kn77} and \citet{kozok85} are listed. As can be seen from
Table~\ref{tab:hist}, in more than 100 years, the mean brightness
of Hen~3-1013 has not changed.

\section{DISCUSSION AND CONCLUSIONS}
\label{sec6}

Based on high-resolution ($R\sim48~000$) observations we have
studied the optical spectrum of the B-supergiant Hen~3-1013,
the central star of the IR-source IRAS~14331-6435. At wavelengths
from 3700 to 8820 \AA, numerous absorption and emission lines
have been identified, their equivalent widths and corresponding
radial velocities have been measured. Using non-LTE model atmospheres,
we have obtained the effective temperature $T_{\text{eff}}=18 250$ K,
gravity $\log g=2.3$ and microturbulence velocity $\xi_{t}=32$ \kms.

The presence of emission lines indicates a low-excitation nebula
surrounding the B-type central star. Nebular expansion velocity
($V_\text{exp}$) is about 12 \kms. It is a typical value for post-AGB
objects.

In the spectrum of Hen~3-1013, unlike the spectra of hotter
post-AGB stars with $T_ {\text{eff}}>20~000$ K, there are as
yet no emission lines ([\ion{N}{II}], [\ion{O}{II}], [\ion{O}{III}],
[\ion{S}{II}]) by which the parameters of the gas shell
($T_{e}$ and $N_{e}$) could be obtained. Analysis of [\ion{Ni}{II}]
lines in the gaseous shell gives an rough estimate for the electron
density $N_e\sim10^7$ cm$^-3$.

It should be noted that the emission line of [\ion{Ni}{II}]
$\lambda$7378 is clearly present, but not identified in the spectra
of the hot post-AGB stars IRAS~18062+2410 \citep{parth00},
IRAS~13266--5551 and IRAS~17311--4924 \citep{sarkar05},
IRAS~22023+5249 \citep{sarkar12}, IRAS~17074--1845, IRAS~17311--4924 and
IRAS~18023--3409 \citep{arkh14}.

Photometric variability of Hen~3-1013 has been detected for the
first time. According to the ASAS data for 2001-2009 and ASAS-SN
data for 2016-2018 the object exhibited brightness variations with
an amplitude of up to 0.2 mag in the $V$ band and a time scale of
several days. As some of us reported previously \citep{arkh07,
arkh12, arkh13, arkh14}, the hot post-AGB candidates also display
fast irregular photometric variability with amplitudes of 0.2--0.4
mag in the $V$ band. For some of them (except for IRAS~19200+3457
and IRAS~19336--0400) there exist spectra of high resolution that
allowed to detect P~Cyg profiles of \ion{He}{I} and \ion{H}{I}
lines indicating mass loss in the stars. In the spectrum
Hen~3-1013 lines \ion{H}{$\alpha$}, \ion{H}{$\beta$} and
\ion{H}{$\gamma$} also have P~Cyg profiles. It was hypothesized
that an unsteady stellar wind is mainly responsible for the
brightness variations.

Published results combined with our new data indicate that
Hen~3-1013 is indeed in the post-AGB phase. Theoretical
calculations of the post-AGB evolution of intermediate-mass stars
predict comparatively short transition times of the star from an
AGB giant to a hot subdwarf and then to a white dwarf. Depending
on the initial mass of the star and the history of mass loss on
the AGB and post-AGB, the HR-diagram crossing time was estimated
to be from 100 to several thousand years \citep{blocker95}. The
modern evolution tracks computed by \citet{bert16} are at least
three to ten times faster.

We estimate the mass of Hen~3-1013 using the recent
post-AGB evolutionary sequences computed by \citet{bert16}.
The core mass $M_{c}\sim$0.58M$_\odot$ for the stellar parameters
derived here ($T_{\text{eff}}$=18~200 K, $\log g$=2.3)
and the metallicity Z(CNO)=0.016 from \citet{mello12} was obtained.
For this mass timescale from the moment in which
$T_{\text{eff}}\sim 7000$ K ($\log T_{\text{eff}}$=3.85)
to $T_{\text{eff}}\sim 18000$ K consist of $\sim$ 100 years.
Since the bolometric luminosity in the post-AGB stage is constant,
the star brightness must track the change of the bolometric
correction with increasing stellar temperature.
According to \citet{fl96}, a temperature rise from 7000 to
18000 K corresponds to a change in the bolometric correction
by 1.7 mag. So we could expect decreasing brightness by 1.7 mag in $V$,
and by 1.3 mag in $B$.

Based on the archival and the new photometric data, we traced the
photometric history of Hen~3-1013 on a timescale longer than
100 years. No significant secular changes in the star brightness
have been found. Thus, we have identified the discrepancies between
observation and new post-AGB models of \citet{bert16}.

% Unfortunate, the parallax of Hen~3-1013 from GAIA DR2: $\pi=-0.14\pm0.14$
% mas cannot be used to derive the distance -- the distance modulus
% -- the luminosity.

\section{ACKNOWLEDGMENTS}

This study was supported by the National Research Foundation (NRF) of
the Republic of South Africa. We wish to thank the administration
of the South African Astronomical Observatory for the allocation
of observing time on the 1.9-m telescope and A.\,Tekola for the
help with observations. A.\,Y.\,K. acknowledges support from
the Russian Science Foundation (project no. 14-50-00043).
This research has made use of the SIMBAD database, operated
at CDS, Strasbourg, France, and SAO/NASA Astrophysics Data System.

\appendix
\newpage
\section{HIGH-RESOLUTION OPTICAL SPECTRUM OF Hen~3-1013}

\begin{figure*}
 \includegraphics[width=1\textwidth, height=1\textheight]{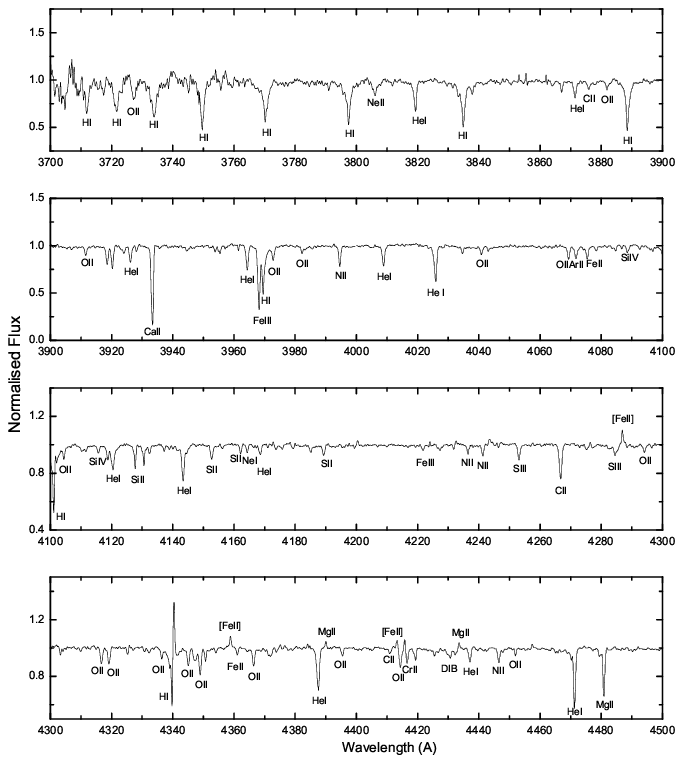}
 \caption{Optical spectrum of Hen~3-1013.}
 \label{fig:sp}
\end{figure*}

\begin{figure*}
 \includegraphics[width=1\textwidth, height=1\textheight]{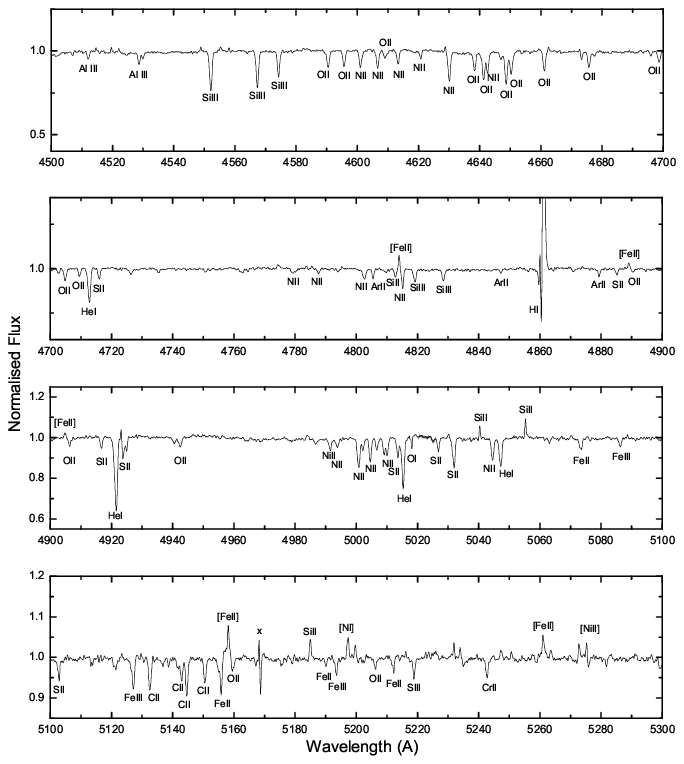}
 \caption{Optical spectrum of Hen~3-1013.}
 \label{fig:sp2}
\end{figure*}

\begin{figure*}
 \includegraphics[width=1\textwidth, height=1\textheight]{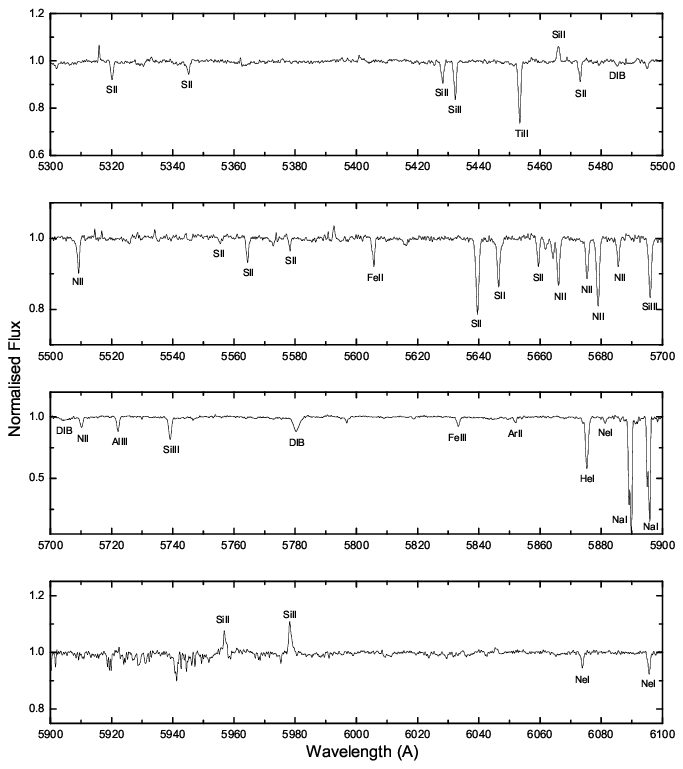}
 \caption{Optical spectrum of Hen~3-1013.}
 \label{fig:sp3}
\end{figure*}

\begin{figure*}
 \includegraphics[width=1\textwidth, height=1\textheight]{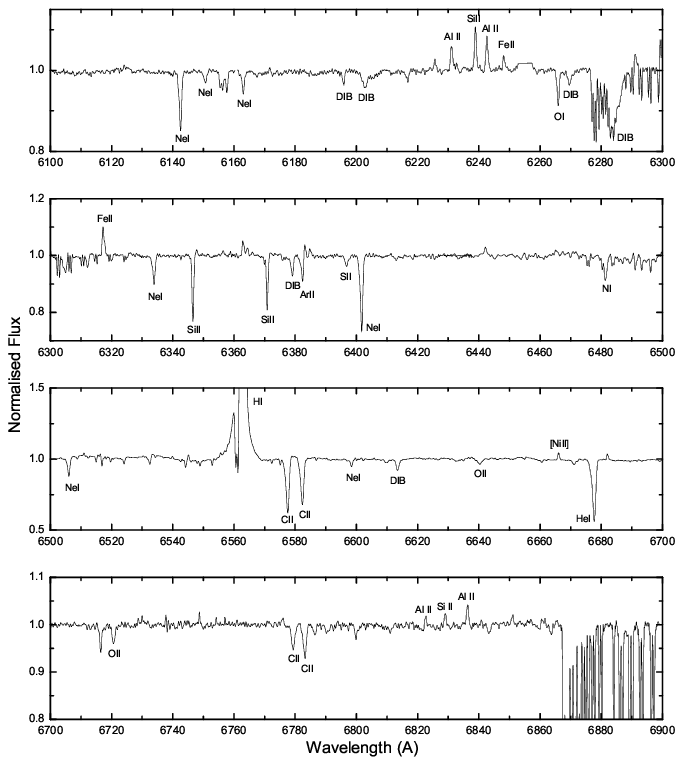}
 \caption{Optical spectrum of Hen~3-1013.}
 \label{fig:sp4}
\end{figure*}

\begin{figure*}
 \includegraphics[width=1\textwidth, height=1\textheight]{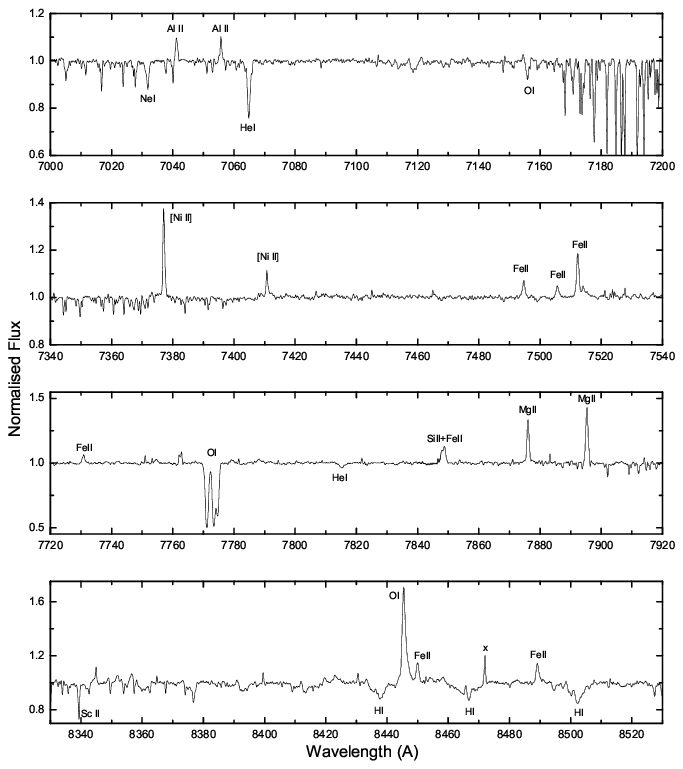}
 \caption{Optical spectrum of Hen~3-1013.}
 \label{fig:sp5}
\end{figure*}

\newpage
\onecolumn
\section{Absorption lines in Hen~3-1013}

\begin{center}
\begin{longtable}{ccccc}
 \caption{Absorption lines in Hen~3-1013}\\
 \label{abslines}
  %\hline

$\lambda_\text{obs.}$& $\lambda_\text{lab.}$&Ident.&
$W_{\lambda}$& $V_\text{r}$\\
(\AA) & (\AA) &&(\AA)& ({\kms}) \\
\hline

\endfirsthead

\multicolumn{2}{l}{continued Table\ref{abslines}}\\
\hline $\lambda_\text{obs.}$& $\lambda_\text{lab.}$&Ident.&
$W_{\lambda}$& $V_\text{r}$\\
(\AA) & (\AA) &&(\AA)& ({\kms}) \\

\hline

\endhead

3717.42&    3717.72&   \ion{S}{III}(6) &0.164  &-24.19\\
3721.65&    3721.94&   \ion{H}{}14 &0.640  &-23.36\\
3733.88&    3734.37&   \ion{H}{}13 &0.796  &-39.34\\
3749.85&    3750.15&   \ion{H}{}12 &0.708  &-23.98\\
3770.15&    3770.63&   \ion{H}{}11 &0.802  &-38.16\\
3797.42&    3797.90&   \ion{H}{}10 &0.609  &-37.89\\
3805.94&    3806.25&   \ion{Ne}{II} &0.217  &-24.42\\
3819.30&    3819.61&   \ion{He}{I}(22)  &0.310  &-24.33\\
3835.00&    3835.40&   \ion{H}{}9  &0.652  &-31.27\\
3837.84&    3838.28&   \ion{S}{III}(5)&0.121  &-34.37\\
3864.00&    3864.43&   \ion{O}{II}(12) &0.054  &-33.36\\
3871.35&    3871.79&   \ion{He}{I}(60) &0.166  &-34.07\\
3875.84&    3876.19&   \ion{C}{II}(33)&0.072  &-27.07\\
3881.90&    3882.20&   \ion{O}{II}(12)&0.078  &-23.17\\
3888.55&    3889.06&   \ion{H}{}8  &0.551  &-39.31\\
3911.56&    3911.96&   \ion{O}{II}(17) &0.072  &-30.65\\
3924.14&    3924.64&   \ion{N}{II}&0.083  &-38.19\\
3926.14&    3926.53&   \ion{He}{I}(58) &0.150  &-29.78\\
3928.14&    3928.54&   \ion{S}{III}(8) &0.042  &-30.52\\
3933.20&    3933.66&   \ion{Ca}{II}(1)&0.745  &-35.06\\
3953.85&    3954.36&   \ion{O}{II}(6) &0.059  &-38.66\\
3955.47&    3955.85&   \ion{N}{II}(6) &0.091  &-28.80\\
3964.32&    3964.73&   \ion{He}{I}(5) &0.238  &-31.00\\
3968.22&    3968.72&   \ion{Fe}{III}(120)&0.635  &-37.77\\
3969.51&    3970.04&   \ion{H}{}7  &0.626  &-40.02\\
3972.80&    3973.26&   \ion{O}{II}(6)&0.164  &-34.71\\
3982.27&    3982.72&   \ion{O}{II}(6) &0.063  &-33.87\\
3994.56&    3995.00&   \ion{N}{II}(12) &0.170  &-33.02\\
4008.97&    4009.51&   \ion{He}{I}(55) &0.190  &-22.43\\
4025.99&    4026.19&   \ion{He}{I}(18) &0.374  &-14.89\\
4040.84&    4041.31&   \ion{O}{II}(50)&0.064  &-34.87\\
4069.38&    4069.89&   \ion{O}{II}(10) &0.131  &-37.57\\
4071.86&    4072.38&   \ion{Ar}{II}(41) &0.148  &-38.28\\
4075.41&    4075.94&   \ion{Fe}{II}(21) &0.136  &-38.98\\
4078.44&    4078.86&   \ion{O}{II}(10)&0.040  &-30.87\\
4084.69&    4085.12&   \ion{O}{II}(10)&0.061  &-31.56\\
4086.81&    4087.27&   \ion{Fe}{II}(28)&0.033  &-33.74\\
4088.50&    4088.86&   \ion{Si}{IV}(19)&0.090  &-26.39\\
4092.56&    4092.93&   \ion{O}{II}(10)&0.034  &-27.10\\
4096.74&    4097.26&   \ion{O}{II}(20)&0.036  &-38.05\\
4101.13&    4101.74&   \ion{H}{}6  &0.379  &-44.58\\
4104.47&    4105.00&   \ion{O}{II}(20)&0.089  &-38.71\\
4115.74&    4116.10&   \ion{Si}{IV}(1) &0.043  &-26.22\\
4118.74&    4119.22&   \ion{O}{II}(20)&0.100  &-34.93\\
4120.42&    4120.82&   \ion{He}{I}(16) &0.247  &-29.10\\
4127.77&    4128.07&   \ion{Si}{II}(3) &0.123  &-21.79\\
4130.58&    4130.89&   \ion{Si}{II}(3) &0.123  &-22.50\\
4132.43&    4132.80&   \ion{O}{II}(19)&0.073  &-26.84\\
4143.42&    4143.76&   \ion{He}{I}(53)&0.287  &-24.60\\
4152.70&    4153.06&   \ion{S}{II}(44)&0.102  &-25.99\\
4162.19&    4162.67&   \ion{S}{II}(44) &0.056  &-34.57\\
4164.37&    4164.81&   \ion{Ne}{I} &0.050  &-31.67\\
4168.61&    4168.97&   \ion{He}{I}(52)&0.060  &-25.89\\
4189.35&    4189.71&   \ion{S}{II}(44)&0.069  &-25.76\\
4221.84&    4222.27&   \ion{Fe}{III} &0.040  &-30.53\\
4227.38&    4227.749&  \ion{N}{II}(33)  &0.056  &-26.26\\
4236.54&    4236.99&   \ion{N}{II}(48) &0.065  &-31.84\\
4241.33&    4241.78&   \ion{N}{II}(48) &0.075  &-31.80\\
4253.20&    4253.50&   \ion{S}{III}(4)&0.099  &-21.14\\
4266.78&    4267.26&    \ion{C}{II}(6) &0.249  &-33.72\\
4284.56&    4284.88&    \ion{S}{III}(4) &0.084  &-22.39\\
4294.14&    4294.40&    \ion{S}{II}(49)&0.042  &-18.15\\
4316.70&    4317.14&    \ion{O}{II}(2)&0.099  &-30.55\\
4319.10&    4319.63&    \ion{O}{II}(2)&0.115  &-36.78\\
4325.29&    4325.76&    \ion{O}{II}(2) &0.032  &-32.57\\
4332.26&    4332.71&    \ion{O}{II}(65)&0.043  &-31.14\\
4336.44&    4336.86&    \ion{O}{II}(2)&0.072  &-29.03\\
4345.11&    4345.56&    \ion{O}{II}(2)&0.105  &-31.04\\
4347.29&    4347.82&    \ion{Al}{II}(70)&0.126  &-36.54\\
4348.93&    4349.43&    \ion{O}{II}(2)&0.167  &-34.46\\
4350.74&    4351.27&    \ion{O}{II}(16)&0.056  &-36.52\\
4354.08&    4354.49&    \ion{S}{III}(7)&0.027  &-28.25\\
4361.05&    4361.59&    \ion{Fe}{II}   &0.048  &-37.12\\
4366.44&    4366.90&    \ion{O}{II}(2)&0.112  &-31.58\\
4387.54&    4387.93&    \ion{He}{I}(51) &0.317  &-26.65\\
4395.53&    4395.95&    \ion{O}{II}(26) &0.046  &-28.64\\
4410.85&    4411.20&    \ion{C}{II}(39)&0.032  &-23.79\\
4414.47&    4414.91&    \ion{O}{II}(5)&0.122  &-29.88\\
4419.33&    4419.78&    \ion{Cr}{II} &0.070  &-30.52\\
4437.15&    4437.55&    \ion{He}{I}(50)&0.087  &-27.02\\
4446.53&    4447.03&    \ion{N}{II}(15) &0.076  &-33.71\\
4451.98&    4452.38&    \ion{O}{II}(5)&0.028  &-26.87\\
4471.17&    4471.48&    \ion{He}{I}(14) &0.327  &-20.78\\
4480.91&    4481.21&    \ion{Mg}{II}(4)&0.255  &-20.07\\
4512.08&    4512.56&    \ion{Al}{III}(3)&0.047  &-31.89\\
4528.57&    4529.16&    \ion{Al}{III}(3)&0.064  &-39.05\\
4552.14&    4552.62&    \ion{Si}{III}(2)&0.246  &-31.61\\
4567.42&    4567.82&    \ion{Si}{III}(2)&0.213  &-26.25\\
4574.25&    4574.76&    \ion{Si}{III}(2)&0.137  &-33.42\\
4590.42&    4590.97&    \ion{O}{II}(15)&0.115  &-35.92\\
4595.76&    4596.17&    \ion{O}{II}(15)&0.099  &-26.74\\
4601.00&    4601.48&    \ion{N}{II}(5) &0.111  &-31.27\\
4606.63&    4607.16&    \ion{N}{II}(15) &0.112  &-34.49\\
4608.99&    4609.44&    \ion{O}{II}(93)&0.075  &-29.27\\
4613.41&    4613.87&    \ion{N}{II}(5)&0.091  &-29.89\\
4620.94&    4621.39&    \ion{N}{II}(5) &0.053  &-29.19\\
4630.03&    4630.54&    \ion{N}{II}(5) &0.173  &-33.02\\
4638.39&    4638.86&    \ion{O}{II}(1)&0.112  &-30.37\\
4641.30&    4641.81&    \ion{O}{II}(1)&0.187  &-32.94\\
4642.66&    4643.08&    \ion{N}{II}(5) &0.125  &-27.12\\
4648.66&    4649.13&    \ion{O}{II}(1)&0.209  &-30.31\\
4650.34&    4650.84&    \ion{O}{II}(1)&0.155  &-32.23\\
4661.15&    4661.64&    \ion{O}{II}(1)&0.119  &-31.51\\
4673.35&    4673.73&    \ion{O}{II}(1)&0.035  &-24.37\\
4675.68&    4676.24&    \ion{O}{II}(1)&0.099  &-35.90\\
4695.96&    4696.36&    \ion{O}{II}(8)&0.016  &-25.53\\
4698.62&    4699.21&    \ion{O}{II}(25)&0.067  &-37.64\\
4704.80&    4705.36&    \ion{O}{II}(25)&0.074  &-35.68\\
4709.45&    4710.00&    \ion{O}{II}(24)&0.046  &-35.01\\
4712.84&    4713.14&    \ion{He}{I}(12) &0.203  &-19.08\\
4715.95&    4716.27&    \ion{S}{II}(9)&0.059  &-20.34\\
4779.11&    4779.71&    \ion{N}{II}(20) &0.044  &-37.63\\
4787.66&    4788.13&    \ion{N}{II}(20) &0.028  &-29.43\\
4802.68&    4803.27&    \ion{N}{II}(20) &0.068  &-36.82\\
4805.43&    4806.02&    \ion{Ar}{II}(6) &0.041  &-36.80\\
4812.82&    4813.33&    \ion{Si}{III}(9)&0.033  &-31.76\\
4815.24&    4815.62&    \ion{N}{II}(20)&0.078  &-23.66\\
4819.26&    4819.72&    \ion{Si}{III}(9)&0.073  &-28.61\\
4828.38&    4828.96&    \ion{Si}{III}(9)&0.063  &-36.01\\
4847.27&    4847.81&    \ion{Ar}{II}(6)&0.019  &-33.39\\
4859.70&    4860.20&    \ion{Cr}{II}(30)&0.047  &-30.84\\
4879.34&    4879.86&    \ion{Ar}{II}(14)&0.047  &-31.95\\
4885.23&    4885.65&    \ion{S}{II}(15)&0.046  &-25.77\\
4890.50&    4890.93&    \ion{O}{II}(28)&0.016  &-26.36\\
4906.42&    4906.83&    \ion{O}{II}(28)&0.047  &-25.05\\
4916.72&    4917.21&    \ion{S}{II}(15)&0.058  &-29.87\\
4921.56&    4921.93&    \ion{He}{I}(48) &0.394  &-22.54\\
4923.69&    4924.12&    \ion{S}{II}(7)&0.096  &-26.18\\
4940.65&    4941.12&    \ion{O}{II}(33)&0.050  &-28.52\\
4986.71&    4987.27&    \ion{Fe}{II} &0.032  &-33.66\\
4991.56&    4991.97&    \ion{S}{II}(7)&0.061  &-24.62\\
4993.83&    4994.36&    \ion{N}{II}(24) &0.054  &-31.81\\
5000.94&    5001.46&    \ion{N}{II}(19) &0.144  &-31.17\\
5002.12&    5002.69&    \ion{N}{II}(4) &0.054  &-34.16\\
5004.67&    5005.14&    \ion{N}{II}(19) &0.102  &-28.15\\
5006.74&    5007.32&    \ion{N}{II}(24) &0.055  &-34.73\\
5009.12&    5009.56&    \ion{S}{II}(7) &0.094  &-26.33\\
5010.03&    5010.62&    \ion{N}{II}(4) &0.084  &-35.30\\
5013.57&    5014.03&    \ion{S}{II}(15) &0.049  &-27.50\\
5015.30&    5015.68&    \ion{He}{I}(4) &0.215  &-22.71\\
5018.21&    5018.78&    \ion{O}{I}(13) &0.027  &-34.05\\
5026.81&    5027.22&    \ion{S}{II}(1)&0.043  &-24.45\\
5031.99&    5032.45&    \ion{S}{II}(7)&0.141  &-27.40\\
5044.47&    5045.10&    \ion{N}{II}(4) &0.100  &-37.44\\
5047.22&    5047.74&    \ion{He}{I}(47) &0.143  &-30.88\\
5073.51&    5074.06&    \ion{Fe}{II}(205)&0.059  &-32.50\\
5086.31&    5086.72&    \ion{Fe}{III}(5) &0.042  &-24.16\\
5102.86&    5103.34&    \ion{S}{II}(7)&0.037  &-28.20\\
5127.12&    5127.50&    \ion{Fe}{III}(5)&0.080  &-22.22\\
5132.58&    5133.10&    \ion{C}{II}(16)&0.079  &-30.37\\
5142.85&    5143.49&    \ion{C}{II}(16) &0.047  &-37.30\\
5144.66&    5145.16&    \ion{C}{II}(16)&0.075  &-29.13\\
5150.39&    5151.08&    \ion{C}{II}(16)&0.055  &-32.01\\
5155.85&    5156.45&    \ion{Fe}{II}   &0.078  &-34.88\\
5159.50&    5159.94&    \ion{O}{II}(32)&0.033  &-25.56\\
5190.19&    5190.74&    \ion{Fe}{II}   &0.013  &-31.77\\
5193.52&    5193.91&    \ion{Fe}{III}(5)&0.029 &-22.51\\
5206.24&    5206.65&    \ion{O}{II}(32)&0.022  &-23.61\\
5212.30&    5212.83&    \ion{Fe}{II}  &0.024  &-30.48\\
5218.81&    5219.32&    \ion{S}{III} &0.039  &-29.29\\
5242.82&    5243.46&    \ion{Cr}{II}(38)&0.053  &-36.59\\
5302.18&    5302.86&    \ion{Si}{III}&0.012  &-38.44\\
5320.20&    5320.73&    \ion{S}{II}(38)&0.075  &-29.86\\
5345.16&    5345.72&    \ion{S}{II}(38)&0.049  &-31.41\\
5428.20&    5428.67&    \ion{S}{II}(6)&0.086  &-25.96\\
5432.31&    5432.82&    \ion{Si}{II}(6)&0.146  &-28.14\\
5453.47&    5454.05&    \ion{Ti}{II}&0.231  &-31.88\\
5473.21&    5473.62&    \ion{S}{II}(6) &0.079  &-22.46\\
5495.03&    5495.67&    \ion{N}{II}(29) &0.036  &-34.91\\
5509.27&    5509.72&    \ion{S}{II}(6)&0.077  &-24.49\\
5555.56&    5556.01&    \ion{S}{II}(6)&0.024 &-24.28\\
5564.49&    5564.94&    \ion{S}{II}(6)&0.068  &-24.24\\
5578.44&    5578.89&    \ion{S}{II}(11)&0.020 &-24.18\\
5605.82&    5606.37&    \ion{Fe}{II} &0.076  &-29.431\\
5639.58&    5639.96&    \ion{S}{II}(14) &0.238  &-20.20\\
5646.55&    5647.02&    \ion{S}{II}(14) &0.141  &-24.95\\
5659.42&    5659.98&    \ion{S}{II}(11) &0.072  &-29.66\\
5661.79&    5662.47&    \ion{C}{II}(15) &0.032  &-36.00\\
5664.31&    5664.77&    \ion{S}{II}(11) &0.059  &-24.34\\
5666.08&    5666.63&    \ion{N}{II}(3) &0.142  &-29.10\\
5675.30&    5676.02&    \ion{N}{II}(3) &0.123  &-38.03\\
5678.99&    5679.56&    \ion{N}{II}(3) &0.218  &-30.09\\
5685.58&    5686.21&    \ion{N}{II}(3) &0.073  &-33.22\\
5695.94&    5696.49&    \ion{Si}{III}&0.181  &-28.95\\
5710.20&    5710.76&    \ion{N}{II}(3) &0.078  &-29.40\\
5722.17&    5722.73&    \ion{Al}{III}(2)&0.111  &-29.34\\
5739.28&    5739.73&    \ion{Si}{III}(4)&0.197  &-23.50\\
5818.69&    5819.27&    \ion{S}{II}(14) &0.018  &-29.36\\
5833.23&    5833.93&    \ion{Fe}{III}(114)&0.087&-35.97\\
5852.14&    5852.74&    \ion{Ar}{II}   &0.034  &-30.73\\
5875.30&    5875.62&    \ion{He}{I}(11) &0.456  &-16.33\\
5881.39&    5881.90&    \ion{Ne}{I}(1)&0.053  &-25.99\\
5927.07&    5927.81&    \ion{N}{II}(28) &0.024  &-37.42\\
5929.04&    5929.69&    \ion{Fe}{III}(114)&0.053  &-32.86\\
5931.16&    5931.78&    \ion{N}{II}(28) &0.034  &-31.33\\
5951.84&    5952.39&    \ion{N}{II}(28) &0.023  &-27.70\\
5975.29&    5975.96&    \ion{Ar}{II}  &0.027  &-33.61\\
6073.84&    6074.34&    \ion{Ne}{I}(3) &0.048  &-24.68\\
6095.65&    6096.16&    \ion{Ne}{I} &0.067  &-25.08\\
6142.52&    6143.06&    \ion{Ne}{I}(1) &0.137  &-26.35\\
6163.04&    6163.59&    \ion{Ne}{I}(5) &0.055  &-26.75\\
6266.09&    6266.89&    \ion{O}{I}(48) &0.061  &-38.27\\
6304.90&    6305.48&    \ion{S}{II}(19) &0.077  &-27.58\\
6312.17&    6312.68&    \ion{S}{II}(26) &0.040  &-24.22\\
6333.91&    6334.43&    \ion{Ne}{I}(1) &0.092  &-24.61\\
6346.64&    6347.10&    \ion{Si}{II}(2)&0.166  &-21.73\\
6370.83&    6371.36&    \ion{Si}{II}(2)&0.128  &-24.94\\
6382.50&    6383.10&    \ion{Ar}{II}  &0.076  &-28.18\\
6396.82&    6397.36&    \ion{S}{II}(19)&0.043  &-25.31\\
6401.78&    6402.25&    \ion{Ne}{I}(1) &0.257  &-22.01\\
6481.17&    6481.71&    \ion{N}{I}(21) &0.094  &-24.98\\
6506.06&    6506.53&    \ion{Ne}{I}(3) &0.117  &-21.66\\
6532.26&    6533.00&    \ion{N}{II}(45)&0.045  &-33.96\\
6577.59&    6578.03&    \ion{C}{II}(2)&0.474  &-20.05\\
6582.33&    6582.88&    \ion{C}{II}(2) &0.405  &-25.05\\
6598.39&    6598.95&    \ion{Ne}{I}(6) &0.033  &-25.44\\
6640.46&    6641.04&    \ion{O}{II}(4)&0.066  &-26.18\\
6677.81&    6678.15&    \ion{He}{I}(46) &0.571  &-15.26\\
6720.66&    6721.38&    \ion{O}{II}(4)&0.044  &-32.11\\
6779.42&    6779.94&    \ion{C}{II}(14) &0.082&-22.99\\
6783.20&    6783.90&    \ion{C}{II}(14) &0.103&-30.93\\
6786.50&    6787.21&    \ion{C}{II}(14) &0.044&-31.36\\
7031.61&    7032.41&    \ion{Ne}{I}(1) &0.116 &-34.10\\
7064.92&    7065.19&    \ion{He}{I}(10) &0.247&-11.46\\
7156.00&    7156.70&    \ion{O}{I}(38) &0.134  &-29.32\\
7280.93&    7281.35&    \ion{He}{I}(45) &0.301  &-17.29\\
7771.15&    7771.94&    \ion{O}{I}(1) &0.670  &-30.32\\
7773.47&    7774.17&    \ion{O}{I}(1) &0.691  &-26.96\\
7774.58&    7775.39&    \ion{O}{I}(1) &0.517  &-31.12\\
7815.23&    7816.10&    \ion{Cr}{II}(69)&0.074 &-33.22\\
7963.44&    7964.15&    \ion{Fe}{II} &0.050  &-26.73\\
8007.78&    8008.65&    \ion{Fe}{II} &0.045  &-32.57\\
8205.07&    8206.04&    \ion{Ne}{II} &0.019  &-35.44\\
8209.94&    8210.53&    \ion{Fe}{II} &0.101  &-21.54\\
8215.69&    8216.34&    \ion{N}{I}(2)&0.033  &-23.72\\
8218.48&    8219.04&    \ion{Fe}{II} &0.140  &-20.44\\
8300.75&    8301.56&    \ion{Ne}{I} &0.123  &-29.25\\
8304.63&    8305.71&    \ion{Fe}{II}&0.083  &-38.98\\
8334.03&    8335.04&    \ion{Sc}{II} &0.029  &-36.18\\
8339.34&    8340.35&    \ion{Sc}{II} &0.142  &-36.30\\
8362.65&    8363.47&    \ion{Al}{II}(40)&0.026  &-29.39\\
8494.52&    8495.36&    \ion{Ne}{I}(18)&0.068  &-29.64\\
8581.89&    8582.61&    \ion{He}{I} &0.295& -25.15\\
8775.85&    8776.71&    \ion{He}{I} &0.215& -29.38\\
8819.51&    8820.43&    \ion{O}{I}(37)&0.150& -31.27\\

 \hline

\end{longtable}
\end{center}


\begin{thebibliography}{99}

%  \bibitem[\protect\citeauthoryear{Arkhipova et~al.} {1999}]{arkh99}
%      Arkhipova  V.P., Ikonnikova N.P., Noskova R.I., Sokol G.V.,
%      Esipov V.F., Klochkova V.G., 1999, Astronomy
%      Letters, 25, 25

 \bibitem[\protect\citeauthoryear{Arkhipova et~al.} {2004}]{arkh04}
     Arkhipova  V.P., Ikonnikova N.P., Komissarova G.V., Noskova R.I., Esipov V.F.,
     2004, Astronomy Letters, 30, 778

 \bibitem[\protect\citeauthoryear{Arkhipova et~al.} {2007}]{arkh07}
     Arkhipova V. P., Esipov V. F., Ikonnikova N. P., Komissarova G. V.,
     Noskova R. I., 2007, Astronomy Letters, 33, 604

 \bibitem[\protect\citeauthoryear{Arkhipova et~al.} {2012}]{arkh12}
     Arkhipova  V.P., Burlak M.A., Esipov V.F., Ikonnikova N.P.,
     Komissarova G.V., 2012, Astronomy Letters, 38, 157

 \bibitem[\protect\citeauthoryear{Arkhipova et~al.} {2013}]{arkh13}
     Arkhipova  V.P., Burlak M.A., Esipov V.F., Ikonnikova N.P.,
     Komissarova G.V., 2013, Astronomy Letters, 39, 619

 \bibitem[\protect\citeauthoryear{Arkhipova et~al.} {2014}]{arkh14}
     Arkhipova V.P., Burlak M.A., Esipov V.F., Ikonnikova N.P.,
     Kniazev A.Yu., Komissarova G.V., Tekola A., 2014, Astronomy
     Letters, 40, 485

%  \bibitem[\protect\citeauthoryear{Asplund et al.}{2009}]{asplund09}
%       Asplund M., Grevesse N., Sauval A.J., Scott P., 2009, ARAA, 47, 481

 \bibitem[\protect\citeauthoryear{Bautista et~al.} {1996}]{baut96}
      Bautista M.A., Peng J., Pradhan A.K., 1996, \apj, 460, 372

 \bibitem[\protect\citeauthoryear{Bl\"{o}cker}{1995}]{blocker95}
      Bl\"{o}cker T., 1995, \aap, 299, 755

 \bibitem[\protect\citeauthoryear{Friedman et~al.} {2011}]
   {fried11} Friedman S. D. et al., 2011, \apj, 727, 33

 %Garcia-Lario et al. 1997

 \bibitem[\protect\citeauthoryear{Garc\'\i a-Hern\'{a}ndez et~al.}{2002}]
 {garcia02} Garc\'\i a-Hern\'{a}ndez D.A., Manchado A., Garc\'\i
 a-Lario P., Domn\'\i guez-Tagle C., Conway G.M., and Prada F.,
 2002, \aap, 387, 955

 \bibitem[\protect\citeauthoryear{Gauba \& Parthasarathy} {2003}] {gp03}
 Gauba G., Parthasarathy M., 2003, \aap,  407, 1007

 \bibitem[\protect\citeauthoryear{Gauba \& Parthasarathy} {2004}] {gp04}
 Gauba G., Parthasarathy M., 2004, \aap, 417, 201

 \bibitem[\protect\citeauthoryear{Gill \& Kapteyn} {1900}]{gill00}
   Gill D., Kapteyn J. C., 1900, Cape Photographic Durchmusterung,
   Part III

 \bibitem[\protect\citeauthoryear{Flower}{1996}] {fl96}
   Flower P.J., 1996, \apj, 469, 355

 \bibitem[\protect\citeauthoryear{Henize} {1976}]{hen76}
   Henize K.G., 1976, \apjs, 30, 491

 \bibitem[\protect\citeauthoryear{Hobbs et~al.} {2008}] {Hobbs08}
   Hobbs L.M., York D.G., Snow T.P., Oka T., Thorburn J.A., Bishof
   M., Friedman S.D., McCall B.J., Rachford B., Sonnentrucker P.,
   Welty D.E., 2008, \apj, 680, 1256

 \bibitem[\protect\citeauthoryear{Hubeny \& Lanz}{1995}]{hubeny95}
   Hubeny I. \& Lanz T., 1995, \apj, 439, 875


 \bibitem[\protect\citeauthoryear{Kaufer et~al.} {1999}] {kauf99}
   Kaufer A., Stahl O., Tubessing S. et al., 1999, The Messenger,
   95, 8

%  \bibitem[\protect\citeauthoryear{Kelly \& Hrivnak}{2005}] {kelly05}
%  Kelly D., Hrivnak B., 2005, \apj 629, 1040

 \bibitem[\protect\citeauthoryear{Klare and Neckel} {1977}] {kn77}
   Klare G., Neckel T., 1977, \aaps, 27, 215

%  \bibitem[\protect\citeauthoryear{Klochkova et~al.} {2002}] {kloch02}
%   Klochkova V. G., Yushkin M. V., Miroshnichenko A. S., Panchuk V. E.,
%   Bjorkman K. S., 2002, \aap, 392, 143

 \bibitem[\protect\citeauthoryear{Kochanek et~al.} {2017}] {koch17}
  Kochanek C.S., Shappee B.J., Stanek K.Z., Holoien T.W.-S.,
  Thompson, Todd A. et~al., 2017, \pasp, 129:104502

 \bibitem[\protect\citeauthoryear{Kozok} {1985}] {kozok85}
  Kozok J.R., 1985, \aaps, 61, 387

 \bibitem[\protect\citeauthoryear{Lanz \& Hubeny}{2007}] {lanz07}
 Lanz T. \& Hubeny I., 2007, \apjs, 169, 83

 \bibitem[\protect\citeauthoryear{Loup et~al.}{1990}] {loup90}
 Loup C., Forveille T., Nyman L. A., Omont A., 1990, \aap, 227, L29

 \bibitem[\protect\citeauthoryear{Lucy}{1995}]{lucy95}
 Lucy L.B., 1995, \aap, 294, 555

 \bibitem[\protect\citeauthoryear{Mello et~al.} {2012}] {mello12}
  Mello  D.R.C., Daflon S., Pereira C.B., and Hubeny I., 2012,
  \aap, 543, A11

 \bibitem[\protect\citeauthoryear{Miller Bertolami}{2016}]{bert16}
     Miller Bertolami M.M., 2016, \aap, 588, A25

 \bibitem[\protect\citeauthoryear{Moore}{1945}] {moore45}
 Moore C. E., 1945, A Multiplet Table of Astrophysical Interest. Princeton
 Univ. Observatory, Princeton

 \bibitem[\protect\citeauthoryear{Nave \& Johansson}
 {2012}]{nave12} Nave G. and Johansson S. (arXiv:1210.4773v1, 2012)

 \bibitem[\protect\citeauthoryear{Parthasarathy \& Pottasch}
 {1989}]{partpot89} Parthasarathy M., Pottasch S.R., 1989, \aap,
 225, 521

 \bibitem[\protect\citeauthoryear{Parthasarathy et~al.} {2000a}]
   {pvd00}  Parthasarathy M., Vijapurkar J., and  Drilling J.S.,
   2000a, \aaps, 145, 269

 \bibitem[\protect\citeauthoryear{Parthasarathy et~al.} {2000b}]{parth00}
   Parthasarathy M., Garc\'\i a-Lario P., Sivarani T., Manchado A., and
   Sanz Fern\'{a}ndez de C\'{o}rdoba L., 2000b, \aap, 357, 241

 \bibitem[\protect\citeauthoryear{Pojmanski} {2002}] {pojm02}
     Pojmanski G., 2002, Acta Astronomica, 52, 397

 \bibitem[\protect\citeauthoryear{Sarkar et~al.} {2005}] {sarkar05} Sarkar
     G., Parthasarathy M., and Reddy B.E., 2005, \aap, 431, 1007

 \bibitem[\protect\citeauthoryear{Sarkar et~al.} {2012}] {sarkar12} Sarkar G.,
    Garc\'\i a-Hern\'{a}ndez D.A., Parthasarathy, Manchado A.,
    Garc\'\i a-Lario P., and Takeda Y., 2012, \mnras, 421, 679

 \bibitem[\protect\citeauthoryear{Shappee et~al.} {2014}] {shap14}
    Shappee B.J., Prieto J.L., Grupe D., Kochanek C.S., Stanek K. Z.,
    De Rosa G., 2014, \apj, 788:48

 \bibitem[\protect\citeauthoryear{Su\'{a}rez et~al.} {2006}]{suarez06}
   Su\'{a}rez O., Garc\'{i}a-Lario P., Manchado A., Manteiga M.,
   Ulla A.  and Pottasch S.R., 2006, \aap, 458, 173

 \bibitem[\protect\citeauthoryear{Otsuka et~al.} {2017}]{otsuka17}
   Otsuka M., Parthasarathy M., Tajitsu A., Hubrig S., 2017, \apj,
   838:71


 \end{thebibliography}
\end{document}